\documentclass[singlecolumn,superscriptaddress]{revtex4}
\usepackage[utf8]{inputenc}
\usepackage{empheq}
\usepackage{amsmath}
\usepackage{amsfonts}
\usepackage{amssymb,bm,txfonts}
\usepackage{mathtools}
\usepackage{stmaryrd}
\usepackage{dsfont}
\usepackage{graphicx}
\usepackage[bookmarksnumbered=true]{hyperref}
\usepackage{stackrel}
\usepackage{color}
\usepackage{comment}
\usepackage{xcolor}
\usepackage{enumitem}
\usepackage{moresize}
\usepackage{graphicx}
\usepackage{siunitx}
\usepackage{fancyhdr}

\def \ee{\end{equation}}
\def \be{\begin{equation}}
\def \eea{\end{eqnarray}}
\def \bea{\begin{eqnarray}}

\begin{document}
\title{Geodesics in Quantum Gravity
}
\author{Benjamin Koch}
\email{benjamin.koch@tuwien.ac.at}
\affiliation{Institut f\"ur Theoretische Physik, Technische Universit\"at Wien, Wiedner Hauptstrasse 8--10, A-1040 Vienna, Austria}
\affiliation{Atominstitut, Technische Universit\"at Wien,  Stadionallee 2, A-1020 Vienna, Austria}
 \affiliation{Pontificia Universidad Cat\'olica de Chile \\ Instituto de F\'isica, Pontificia Universidad Cat\'olica de Chile, \\
Casilla 306, Santiago, Chile}
\author{Ali Riahinia}
\email{ali.riahinia@tuwien.ac.at}
\affiliation{Institut f\"ur Theoretische Physik, Technische Universit\"at Wien, Wiedner Hauptstrasse 8--10, A-1040 Vienna, Austria}
\affiliation{Atominstitut, Technische Universit\"at Wien,  Stadionallee 2, A-1020 Vienna, Austria}
\author{Angel Rincon}
\email{angel.rincon@physics.slu.cz}
\affiliation{
Research Centre for Theoretical Physics and Astrophysics, Institute of Physics, Silesian University in Opava,
Bezručovo náměstí 13, CZ-74601 Opava, Czech Republic.
}
%
\begin{abstract}
We investigate the motion of test particles in quantum-gravitational backgrounds by introducing the concept of q--desics, quantum-corrected analogs of classical geodesics. Unlike standard approaches that rely solely on the expectation value of the spacetime metric, our formulation is based on the expectation value of quantum operators, such as the the affine connection-operator. This allows us to capture richer geometric information. We derive the q--desic equation using both Lagrangian and Hamiltonian methods and apply it to spherically symmetric static backgrounds obtained from canonical quantum gravity.
Exemplary results include, light-like radial motion
and circular motion with quantum gravitational corrections far above the Planck scale.

This framework provides a refined description of motion in quantum spacetimes and opens new directions for probing the interface between quantum gravity and classical general relativity.
\end{abstract}
\maketitle
\tableofcontents
\section{Introduction}

The concept of geodesics is utterly fundamental to the framework of General Relativity (GR) and its profound impact on our understanding of gravity and the cosmos. From the earliest triumphs of the theory, such as the explanation of Mercury's anomalous perihelion precession \cite{Einstein:1916vd} and the celebrated confirmation of light deflection by the Sun during the 1919 Eddington expedition \cite{Dyson:1920cwa}, geodesics, the paths taken by free-falling particles and light rays, have been the primary means by which GR makes contact with observation. Indeed, a vast array of astrophysical and cosmological phenomena, including gravitational lensing \cite{Bartelmann:1999yn}, the dynamics of celestial bodies, the Shapiro time delay \cite{Shapiro:1964uw}, and even the propagation paths of gravitational waves~\cite{LIGOScientific:2016aoc}, are interpreted and predicted through the calculation of timelike or null geodesics within curved spacetime. Consequently, the geodesic equation serves as an indispensable tool, forming the bedrock for virtually every observable test and application of Einstein's theory.

Despite its remarkable successes, GR is widely regarded as a classical theory, expected to break down when approaching the quantum regime. The need to reconcile GR with the principles of quantum mechanics (QM) and to achieve a consistent description of gravitational interactions at all scales has spurred the development of various approaches towards a theory of Quantum Gravity (QG). Prominent among these endeavors are 
Canonical QG~\cite{Arnowitt:1959ah,Arnowitt:1962hi},
String Theory~\cite{Aharony:1999ti}, Loop Quantum Gravity~\cite{Rovelli:1994ge}, Asymptotic Safety~\cite{Reuter:2012id}, Causal Dynamical Triangulations~\cite{Ambjorn:2012jv}, and other frameworks. While differing in their foundational assumptions and mathematical machinery, these theories collectively anticipate that the classical description of spacetime geometry emerges as an approximation from a more fundamental, quantum-mechanical substrate.

An undeniably important point to mention at this stage is that, the study of how particles evolve in curved spacetime is a seminal aspect of a well-defined classical gravitational theory. Once a consistent theoretical formulation of QG has been established (e.g. within of one of the afore mentioned approaches), it only can become a physical theory once it allows for predictions, which are in principle testable. Thus, 
we need to understand the concept of observables such as geodesics within any more fundamental framework such as QG. This is the reason for this research.\\
In particular,
given the pivotal role of geodesics in classical GR (as highlighted in the first paragraph) and the expectation that GR itself is a limiting case of a more comprehensive QG theory (as discussed in the second), a critical re-evaluation of how particle trajectories are described within a quantum-gravitational context is imperative. A common first step involves considering particle motion along geodesics of an effective background metric, wherein quantum corrections are absorbed into a modified classical geometry, often viewed as the expectation value of some metric operator. However, it is pertinent to question whether such a semi-classical approximation fully captures the intricacies of motion in a quantum spacetime. As the affine connection and the metric tensor govern the propagation of test particles in GR, studying their quantum counterparts suggests that we may need to go beyond a mere modification of the background. The influence of quantum fluctuations, potential non-local effects, or interactions involving higher-order combinations of metric operators acting on the quantum state of spacetime could lead to a richer, more complex notion of particle paths. This paper aims to explore such a deeper framework, leading to the development and analysis of what we term ``q--desics''--equation of motion that endeavors to encapsulate these more fundamental quantum-gravitational influences on particle trajectories.

\subsection{The Geodesic Equation
in Classical Backgrounds}

A cornerstone of classical General Relativity (GR) is the geodesic equation, which governs the free-fall motion of test particles in curved spacetime. Historically, this insight dates back to Einstein's formulation of GR, wherein the equivalence principle provides a guiding conceptual foundation: inertial motion in a gravitational field is interpreted as motion along a geodesic of the classical spacetime manifold.

In this subsection, both the spacetime metric $g_{\mu\nu}$ and the point particle itself are treated as purely classical entities; no quantum considerations are introduced here. 
%
%
In modern treatments, the geodesic equation is often derived from an action principle for a classical point particle of mass $m$ moving in a background metric $g_{\mu\nu}$. The relevant action,
\begin{equation}\label{eq_SRPP0}
S[g_{\mu \nu}] \;=\;\int d\lambda \,\sqrt{\,g_{\mu \nu} \;\frac{dx^\mu}{d\lambda}\,\frac{dx^\nu}{d\lambda}}\,,
\end{equation}
can be viewed as the proper length (or proper time, if one wishes to pick a specific parametrization) traversed by the particle between two events in a purely classical spacetime. Varying this action with respect to the particle trajectory $x^\mu(\lambda)$ yields the geodesic equation:
\begin{equation}\label{eq_geodcl}
\frac{d^2 x^\mu}{d\lambda^2}
\;+\;\Gamma^\mu_{\alpha\beta}
\,\frac{dx^\alpha}{d\lambda}\,\frac{dx^\beta}{d\lambda}
\;=\;0\,,
\end{equation}
where $\Gamma^\mu_{\alpha\beta}$ denotes the Christoffel symbols associated with the classical metric $g_{\mu\nu}$.

The geodesic equation thus asserts that a free particle experiences no external force in the usual Newtonian sense; rather, its worldline is ``bent'' by the local curvature of the spacetime manifold. Physically, geodesics describe how freely falling particles move under the influence of gravity alone, making them a central tool in analyzing motion near massive bodies or in cosmological backgrounds. Historically and conceptually, establishing the geodesic principle (i.e.\ ``free bodies follow geodesics'') as a natural consequence of the classical field equations of GR was a major milestone in understanding gravity as geometry.
The geodesic equation governs the motion of free particles in curved spacetime. Its simplest formulation arises in general relativity. Consequently, it does not account for quantum effects at any scale.

Moreover, as mentioned in the previous subsection, nearly every observation in modern cosmology and astrophysics---from the orbital motion of stars in galaxies, to the lensing of light by massive clusters, to the redshift--distance relation in an expanding Universe---relies on the classical geodesic equation. Any potential deviation from this law in the presence of new physics (quantum or otherwise) could dramatically alter the predictions for astrophysical and cosmological observations. Consequently, understanding the classical derivation of the geodesic equation and recognizing where one might expect departures from it is essential for comparing QG-theory to data and, eventually, for probing potential quantum-gravitational effects.

\subsection{From Quantum Gravity to Geodesic Motion}

\begin{figure}
    \includegraphics[width=0.80\linewidth]{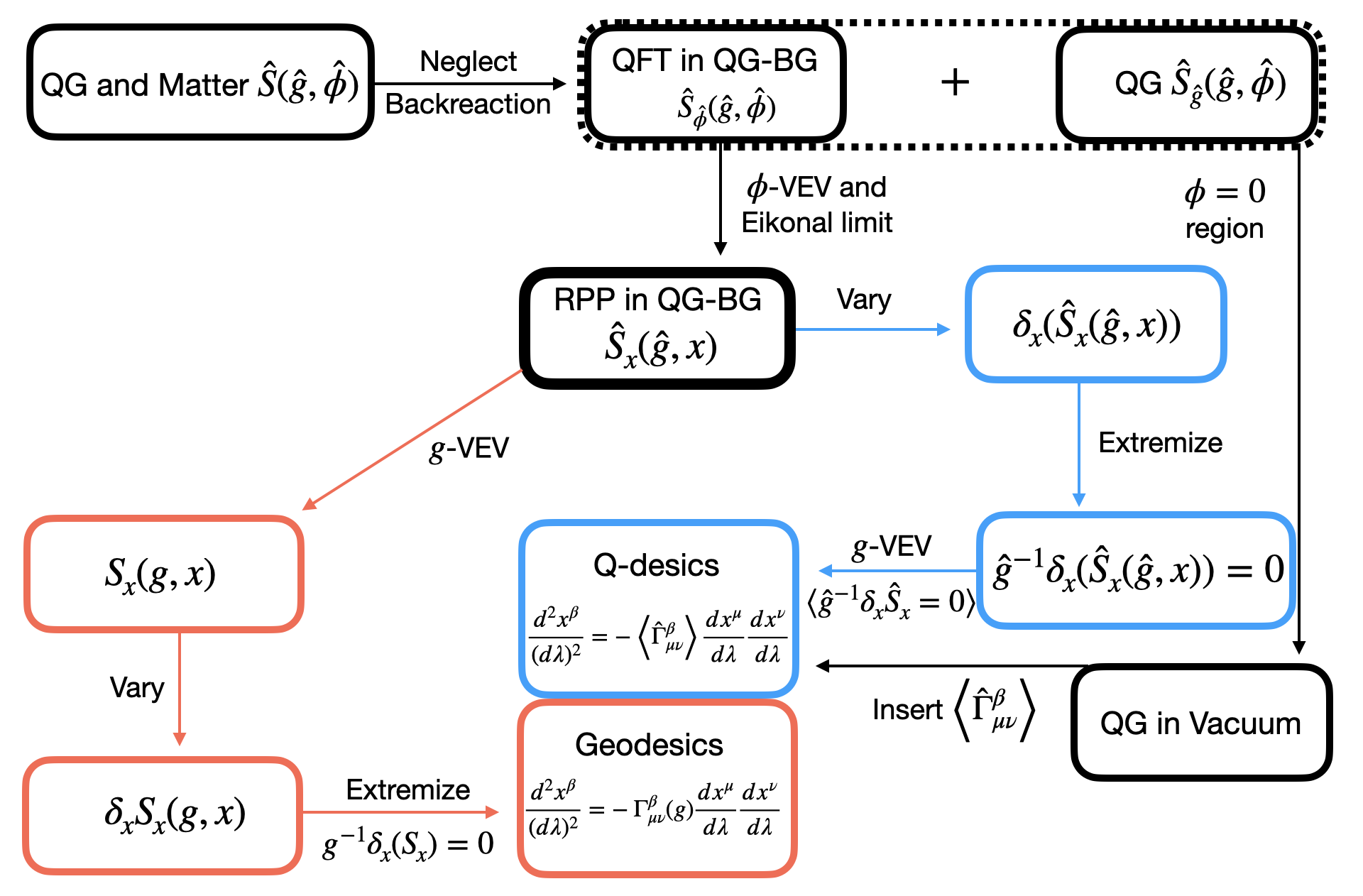}
    \caption{Conceptual path to geodesic equation(s).  
    Hatted quantities denote quantum operators. Along the conventional red path, all vacuum expectation values (VEVs) are taken before variation and extremization, yielding the standard geodesic equation.  
    In contrast, the blue path defers the VEV of the metric degrees of freedom until the final step. This leads to a quantum-modified equation of motion, which we refer to as the q--desic equation.  
    Solving this equation requires input from quantum gravity in the form of operator averages such as $\left\langle \hat{\Gamma}^\beta_{\mu \nu} \right\rangle$.}
    \label{fig:PTG}
\end{figure}

Imagine we have a consistent theory of quantum gravity coupled to matter. How would the geodesic equation emerge from such a theory?  
Even though we do not currently possess this complete theory, we can envision a conceptual roadmap outlining the steps required to arrive at classical geodesic motion.  
This roadmap is illustrated in Figure~\ref{fig:PTG}.

We can trace this conceptual chart from top to bottom. The starting point is a hypothetical but well-defined quantum theory of both metric operators $\hat{g}$ and matter fields $\hat{\phi}$.  

We first simplify by neglecting the backreaction of matter on the metric and vice versa. 
It is often justifiable to neglect back-reaction in quantum gravity (or in theories that incorporate quantum effects) in certain contexts; whether this is appropriate depends on the physical system and the regime under study. Roughly speaking, back-reaction refers to the influence of quantum fields on the background-spacetime geometry. In classical general relativity, matter shapes spacetime through the Einstein field equations. In quantum gravity, however, quantum fluctuations of matter fields could, in principle, modify the spacetime metric.
In studies of geodesic motion, back-reaction is typically neglected because the framework treats the moving objects as test particles that follow geodesics in a fixed spacetime and exert a negligible influence on the geometry itself.
This results in two decoupled quantum theories: one for matter, $\hat{S}_{\hat\phi}$, with a fixed quantum geometry, and one for geometry, $\hat{S}_{\hat g}$, with matter acting as background.

From the matter sector, we may take the vacuum expectation value and the eikonal limit of the quantum field $\hat{\phi}$. This leads to an effective point-particle action in a quantum-gravitational background described by metric operators $\hat{g}$. This intermediate theory is labeled in Figure~\ref{fig:PTG} as the Relativistic Point Particle in Quantum-Gravitational Background (RPP in QG-BG), shown in the central black box.\\
To derive the standard geodesic equation from this point-particle theory, we perform the following three steps:
\begin{enumerate}
    \item Take the expectation value of the metric operators: $\hat{g} \rightarrow \langle \hat{g} \rangle=g$,
    \item Vary the action with respect to the particle trajectory: $\delta_x S$,
    \item Extremize by imposing the stationary action condition after multiplying with the inverse metric: $g^{-1}\delta_x S = 0$.
\end{enumerate}
The central idea of this letter is to point out that the order of these three operations matters.  
In Figure~\ref{fig:PTG}, this is illustrated 
by contrasting two distinct routes:
\begin{itemize}
    \item {\textcolor{red}{The red path }} follows the standard sequence: 1~$\rightarrow$~2~$\rightarrow$~3, resulting in the classical geodesic equation.
    \item {\textcolor{blue} {The blue sequence }}changes the order: 2~$\rightarrow$~3~$\rightarrow$~1. It leads to a more general equation of motion -- the q--desic equation.
\end{itemize}
In this q--desic framework, quantum aspects of spacetime manifest in the appearance of averaged geometric structures, particularly the vacuum expectation values of the Christoffel symbols
$
\left\langle \hat{\Gamma}^\beta_{\mu \nu} \right\rangle.$
These are obtained from the underlying quantum gravity theory of the metric, evaluated in a background where matter is absent or plays the role of a background.

Finally, while we disregard the back-reaction effect intentionally, our approach treats the metric as an operator. This gives us equations for gravitons coupled to a classical source. Further details can be found in references \cite{Skagerstam:2018jkw} (general relativity) and \cite{Skagerstam:2018ofu} (electromagnetism). In the latter case, various forms of classical motion lead to states outside Fock space, and similar reasoning applies to gravity (see \cite{dimock2021quantumradiationclassicalpoint}).
Even more, a path metric path integral might even induce a quantization of matter degrees of freedom~\cite{Koch:2023dfz} or spacetime superposition~\cite{Foo:2022dnz,Foo:2023vbr,Suryaatmadja:2023onb}.
On these grounds, the self-consistency of the proposed formulation as an approximation to a complete quantum gravity (QG) path integral framework is open to discussion. 
Still, our approach provides a controlled generalization of classical geodesics since the metric operator, depending on how it is calculated, can contain all aspects of quantum gravity. However, in the concrete examples that will be provided in Section III, the properties of the metric operator is limited to approximations such as spherical symmetry, or the absence of virtual mater states  and and higher curvature contributions. Thus, it is evident that our formalism successfully captures quantum features, whereas a full fledged QG formulation neither anticipated
nor necessary at the current stage.

\subsection{Working Hypothesis and Structure of the Paper}

We hypothesize that the motion of test particles in a quantum-gravitational background is not governed by the classical geodesic equation in terms of the expectation value metric $g_{\mu \nu}=\langle \hat{g}_{\mu\nu} \rangle$, but rather by an operator-averaged equation of motion in which the variation is taken before expectation values are applied.

As it will be shown, this leads to the proposal that particle trajectories obey a modified geodesic equation of the form:
\[
\frac{d^2 x^\mu}{d \lambda^2} + \left\langle \hat{\Gamma}^\mu_{\nu \rho} \right\rangle \frac{dx^\nu}{d\lambda} \frac{dx^\rho}{d\lambda} = 0,
\]
where the connection coefficients are given by the quantum expectation value of the Christoffel operator, rather than those computed from the classical background metric.

We refer to solutions of this equation as q--desics, quantum-corrected analogues of classical geodesics. These trajectories incorporate effective quantum gravity corrections to motion without requiring a classical spacetime background, and provide a testable link between low-energy observables and the operator structure of a fundamental quantum gravity theory.
Our primary goal is to establish that if quantum effects are taken into account from the very beginning, the equations governing even the motion of classical particles are significantly modified. 
Our results represent an essential step toward formulating a self-consistent framework in which general relativity and quantum mechanics can coexist, and in which the distinction between them can lead to observable consequences.

The structure of the article is as follows:

In Section~I, we introduce the problem of geodesic motion in classical and quantum-gravitational backgrounds, motivate the q--desic framework, and present our working hypotheses. 

Section~II develops the geodesic equation in a quantum-gravitational setting, offering derivations from both Lagrangian and Hamiltonian formalisms and introducing the concept of the q--desic. 

Section~III focuses on spherically symmetric static backgrounds, reviewing relevant aspects of quantum gravity in this context and constructing the necessary expectation values.

Section~IV applies the q--desic formalism to this class of backgrounds, analyzing the resulting motion in two complementary regimes, including radial (null)-motion and  circular orbits. 

Section~V provides a broader discussion of conceptual aspects, such as the role of quantum averages and connections to the effective average action approach, as well as the interpretation of integration constants arising in the formalism. This section concludes with a discussion of novel observable effects at small and large distance scales.

Finally, Section~VI summarizes the main findings.

\section{Geodesic Equation in Quantum-Gravitational Backgrounds}

Our aim is to derive a generalized
geodesic equation which is also valid
in backgrounds with quantum aspects.
The starting point of this discussion is 
the action of a relativistic point particle 
with the velocity $dx^\mu/ d\lambda$
coupling to a gravitational background $g_{\mu \nu}$ is given in (\ref{eq_SRPP0})
where we are using the metric signature $(+,-,-,-)$.
In this action, both the particle and the metric are quantum objects. The most commonly used quantization approaches the canonical formalism and the path integral formalism. This means that we have to treat $g_{\mu \nu}$ and $x^\mu$ either as operators in a canonically quantization approach, or as fields in a path integral picture. We opt to treat the metric canonically $g_{\mu \nu}\rightarrow \hat g_{\mu \nu}$ and the particle as field
\be\label{eq_SRPP1}
\hat S(x^\mu, \hat g_{\mu \nu})=\int d \lambda \sqrt{\hat g_{\mu \nu}\frac{dx^\mu}{d\lambda}\frac{dx^\nu}{d\lambda}}.
\ee
To arrive at the concepts of geodesics and q--desics, we need to make certain assumptions and define some operations:
\begin{itemize}
    \item Operator products\\
    For given tensorial operators $\hat T^\alpha_{\dots}, \hat W_{\beta}^{\dots}$, we need to define products and contractions. Since these operators do not necessarily commute we use Weyl ordered products for products that are observables. For example, we would write an observable
    \be
\hat T^\alpha_{\dots} \hat W_{\beta}^{\dots}\equiv 
    \frac{1}{2}\left(\hat T^\alpha_{\dots} \hat W_{\beta}^{\dots}    +\hat W_{\beta}^{\dots}\hat T^\alpha_{\dots} \right).
    \ee
 This ordering ensures a Hermiticity of products of hermitian operators in the associated observables. Note that this ordering prescription only applies to the outer layer of a product as it is written in an observable and not to deeper layers e.g. using further commutators $(i \hbar \hat 1)_W =i \hbar \hat 1\neq ([\hat x, \hat p])_W=0$.
    \item Average\\
    An expectation value with respect to the wave function $|\Psi\rangle$ of the spacetime operator shall be denoted as
    \be\label{eq_VEVg}
    \langle \Psi |\hat g_{\mu \nu}|\Psi \rangle =\langle \hat g_{\mu \nu} \rangle_g.
    \ee
    \item Inverse metric operator\\
    We assume that the inverse metric operator exists such that 
    \be\label{eq_ginv}
    \hat g_{\mu \nu}\hat g^{\nu \rho}=\delta_\mu^{\; \rho}\hat 1.
    \ee
    Note that this is a non-trivial assumption, as it will be discussed in subsection~\ref{subsec_Comments}.
    \item Variations\\
    Variations with respect to the metric $\delta_g$ shall not concern us in this work, since we consider (quantum)-spacetime as a background without backreaction. Covariant variations with respect to the position of the test particle shall be labeled 
    \be
    \delta_x (x^\nu) = \delta_x x^\nu.
    \ee
    Since the position dictates, where the action is evaluated, these variations also induce variations on the metric
    \be
    \delta_x \hat g_{\mu \nu}=\hat g_{\mu \nu, \alpha}\delta_x x^{\alpha}.
    \ee
\end{itemize}

\subsection{Derivation from a Lagrangian Action}

The action (\ref{eq_SRPP0}) has reparametrization invariance as gauge symmetry. To avoid seeming ambiguities associated to this gauge invariance, we fix the gauge. As shown in~\cite{Rizzuti:2019nes,Gueorguiev:2019zvq}, it is convenient to choose a gauge fixing where the action is linear in the metric
\be\label{eq_SRPP1b}
 \hat S(x^\mu, \hat g_{\mu \nu})=\int d \lambda\; \hat g_{\mu \nu}\frac{dx^\mu}{d\lambda}\frac{dx^\nu}{d\lambda}.
\ee
In this operator valued action, we already took the average with respect to the particle wave function.

In a quantum description, we are used to the fact that two operations do not commute (e.g. $\langle | \hat p^2|\rangle\neq \langle | \hat p|\rangle^2 $). The same is true for the variations and averages introduced above.
The conventional concept of geodesics can be derived from (\ref{eq_SRPP1b}) by first taking both particle and spacetime averages and then vary with respect to the position $\delta_x \langle  \langle \hat S \rangle_x \rangle_g$, as shown by the red path in figure \ref{fig:PTG}. 
In doing so, one looses track of eventual quantum properties of the metric background.
We will now explore how this concept can be generalized if we change that order as indicated by the blue path in figure \ref{fig:PTG}.
First we extremize the operator relation with respect to this position and only finally we calculate the expectation value of the spacetime wave function:
\bea
 \delta_x \hat S(x^\mu, \hat g_{\mu \nu})  &=&
\int d \lambda  \; \delta_{ x} \left(
\hat g_{\mu \nu}\frac{d x^\mu}{d\lambda}\frac{d x^\nu}{d\lambda}
\right)\\
&=&
\int d \lambda  \left(
\left( \delta_{ x}\hat g_{\mu \nu} \right) \frac{d x^\mu}{d\lambda}\frac{d x^\nu}{d\lambda} + \hat g_{\mu \nu} \left( \delta_{ x}\frac{d x^\mu}{d\lambda} \right) \frac{d x^\nu}{d\lambda}
+ \hat g_{\mu \nu}\frac{d x^\mu}{d\lambda} \left( \delta_{ x}\frac{d x^\nu}{d\lambda} \right) \right)\\
&=&
\int d \lambda \left( \hat g_{\mu \nu,\alpha} \delta_{ x}  x^{\alpha} \; \frac{d x^\mu}{d\lambda}\frac{d x^\nu}{d\lambda}\; + 2 \hat g_{\mu \nu}\frac{d x^\mu}{d\lambda}\frac{d \delta_{ x} x^\nu}{d\lambda}\right).
\eea
Next, we integrate by parts in the second term and drop a boundary term. This yields
\bea
\delta_x \hat S(x^\mu, \hat g_{\mu \nu})
&=&
\int d \lambda \left(
(\delta_x x^{\alpha}) \hat g_{\mu \nu , \alpha}  \frac{dx^{\mu}}{d\lambda} \frac{dx^{\nu}}{d\lambda} -
2 (\delta_x x^{\alpha}) \hat g_{\mu \alpha} \frac{d^2 x^{\mu}}{d\lambda^2} - 
2 (\delta_x x^{\alpha}) \frac{d x^{\mu}}{d \lambda} \frac{d x^{\nu}}{d \lambda} \hat g_{\mu \alpha , \nu}
\right).
\eea
The contravariant variations $\delta_{ x}  x^\alpha$ can be pulled out, giving
\bea
\delta_x \hat S(x^\mu, \hat g_{\mu \nu})
&=&
\int d \lambda \; 2\left[
 \frac{d x^\mu}{d\lambda}\frac{d x^\nu}{d\lambda}
 \left( \frac{1}{2}
\left(\hat g_{\mu \nu,\alpha}
\; 
-\hat g_{\mu \alpha,\nu}-\hat g_{\nu \alpha,\mu} \right)\right)
-
\frac{d^2 x^\mu}{d\lambda^2} \hat g_{\mu \alpha}
 \right] \delta_{ x} x^\alpha.
\eea
Now, by imposing that this action is extremal for arbitrary variations $\delta_{ x} x^\alpha$, we can read 
\be\label{eq_geodOp}
\frac{d x^\mu}{d\lambda}\frac{d x^\nu}{d\lambda}
\left( \frac{1}{2}
\left(\hat g_{\mu \alpha,\nu} + \hat g_{\nu \alpha,\mu} -\hat g_{\mu \nu,\alpha} \right) \right) + \frac{d^2 x^\mu}{d\lambda^2} \hat g_{\mu \alpha} = 0.
\ee
This operator valued identity contains all the information the variational principle can provide us.
To reduce it to a simple differential equation, we need to take the VEV with respect to the metric degrees of freedom~(\ref{eq_VEVg}).
We realize that if we perform this operation on 
(\ref{eq_geodOp}) in its current form and multiply the result with the VEV of the inverse metric $g^{\alpha \beta}$, we recover the usual geodesic equation (\ref{eq_geodcl}).
Inverting the order between variation and averaging had no effect on the resulting differential equation, so it seems.

However, there is a subtlety in the final step of the procedure: Whenever we take the VEV of a non-trivial quantum operator, a projection takes place and information is lost. For example, when we take the metric VEV of equation (\ref{eq_geodOp}), we loose  information about wave functions $| \Psi\rangle$ which are not in an eigenstate of the metric operator $\hat g_{\mu \nu}$, or in an eigenstate of the operator $
\left(\hat g_{\mu \alpha,\nu} + \hat g_{\nu \alpha,\mu} -\hat g_{\mu \nu,\alpha} \right)$.
Clearly, we would like to reduce this loss of information to a minimum, e.g. by making one side of the equation proportional to the unit operator $\hat 1$. Since $\langle \Psi | \hat 1 | \Psi \rangle=1$, 
the wave function would be projected to itself and less information would be lost. Indeed, 
the right hand side of (\ref{eq_geodOp}) can be made proportional to the unit operator $\hat 1$ by invoking (\ref{eq_ginv}) and multiplying (\ref{eq_geodOp}) with $\hat g^{\alpha \beta }$.
Finally, taking the expectation value
with respect to the metric degrees of freedom gives the anticipated q--desic equation
\be\label{eq_Geodesic1}
\frac{d^2 x^\beta}{d\lambda^2} +
 \left \langle \hat \Gamma^\beta_{\mu \nu}\right \rangle_g
 \frac{d x^\mu}{d\lambda}\frac{d x^\nu}{d\lambda}=0,
\ee
where we defined
\be\label{eq_GammaDef}
\left \langle \hat \Gamma^\beta_{\mu \nu}\right \rangle_g \equiv  \left \langle \frac{1}{2}
\left(
\hat g_{\mu \alpha,\nu}+\hat g_{\nu \alpha,\mu} -\hat g_{\mu \nu,\alpha}\right)
\hat g^{\alpha \beta}\right\rangle_g.
\ee
In what follows
we 
will dropped the index $_g$, since it is clear that the only remaining expectation value is the one with respect to the wave function of the metric operator.
The procedure shown in this subsection was largely simplified by our choice of gauge in the Lagrangian action.

\subsection{Derivation from a Hamiltonian Action}

If we prefer  to postpone fixing the gauge degree of freedom for the relativistic point particle, we can also work with the explicitly gauge invariant Hamiltonian action
\be\label{eq_SH}
\hat S_H(x^\mu, p^\mu,\hat g_{\mu \nu})= \int d\lambda \left(p^\mu \hat g_{\mu \nu}\dot{ x}^\nu-\frac{\lambda}{2} \left(p^\mu \hat g_{\mu \nu}p^\nu-m^2 \ \right)\right).
\ee
Alternatively, we could also define all $x^\mu$ contravariant and all momenta covariant. This does not alter the final result.
Now we vary this operator valued action with respect to the constraint $\delta_\lambda$ and the particle variables
$\delta_x\equiv \delta x^\nu \partial_\nu$,
$\delta_p\equiv\delta p^\nu \partial/(\partial p^\nu)$.
Extremizing the variation with respect to the constraint gives
\be\label{eq_constraint}
\left(p^\mu \hat g_{\mu \nu}p^\nu-m^2 \ \right)=0.
\ee
Extremizing  with respect to the momentum gives
\be\label{eq_deltap}
(\dot x^\nu - \lambda p^\nu)\hat g_{\mu \nu}\delta p^\mu=0. 
\ee
We realize that the condition (\ref{eq_deltap}) can be fulfilled if 
\be\label{eq_deltapweak}
(\dot x^\nu - \lambda p^\nu)=0
\ee
We extremize (\ref{eq_SH})
with respect to $\delta x$, and obtain after a partial integration
\be\label{eq_deltax1}
\left[\dot p^\mu \hat g_{\mu \nu}+\lambda p^\mu p^\alpha \hat g_{\mu \nu,\alpha}
- \frac{\lambda}{2}p^\alpha p^\beta g_{\alpha \beta,\nu}\right]\delta x^\nu=0.
\ee
Now, we first multiply by $\lambda$ and use the condition (\ref{eq_deltapweak}) and partially fix the gauge such that $\dot \lambda=0$.
After this, we find
\be\label{eq_deltax2}
\left[ \ddot x^\mu \hat g_{\mu \nu}+
\frac{1}{2}\dot x^\alpha \dot x^\beta \left(
\hat g_{\nu \beta,\alpha}+ \hat g_{\nu \alpha, \beta}-\hat g_{\alpha \beta, \nu}
\right)\right]\delta x^\nu=0.
\ee
Taking the metric VEV of this relation would again provide the geodesic equation (\ref{eq_geodcl})
and lead to a potential loss of information about the spacetime state $|\Psi\rangle$.
To avoid this loss, invoke (\ref{eq_ginv})
and insert the operator identity $\hat \delta^\gamma_\nu=\hat g_{\nu \mu} \hat g^{\mu \gamma}$ in the second term of these brackets. We can factorize a metric operator
and read off the operator valued equation 
\be\label{eq_deltaxweak}
\left[ \ddot x^\mu +
\dot x^\alpha \dot x^\beta \hat \Gamma^\mu_{\; \alpha \beta}\right]=0.
\ee
In the final step we take the expectation value of this relation and recover the q--desic equation (\ref{eq_Geodesic1}).

The q--desic equation (\ref{eq_Geodesic1}),
combined with the expectation value of the mass constraint relation (\ref{eq_constraint}),
while the latter can be conveniently written as
\be\label{eq_properTimeu2}
\langle \hat g_{\mu \nu}\rangle u^\mu u^\nu =\left\{ 
\begin{array}{cc}
     1& {\text{for massive particles}} \\
     0 & {\text{for massless particles,}}
\end{array}
\right.
\ee
Before we proceed and solve the equations (\ref{eq_Geodesic1} and \ref{eq_properTimeu2}),
we need a method to actually calculate the operator averages appearing in both equations.
Such a method will be revisited in the following section.

\section{Quantum Gravity of Spherically Symmetric Static Backgrounds}
\label{sec_SER}

A gravitational system of particular simplicity and interest is a static spacetime with spherical symmetry. 
The representative hypersurfaces of this type of spacetime is characterized by a Static Equal Radius (SER) condition.

\subsection{Static Spherically Symmetric Quantum Gravity in a Nutshell}

In this subsection we will shortly summarize ingredients that are necessary for the calculation of $\langle \hat \Gamma^\alpha_{\; \mu \nu} \rangle$
in static spherically symmetric quantum gravity. 
Details, derivations, and discussions on this topic are given in~\cite{Koch:2025yuz}.
Note that also other approaches to quantum-gravitational spacetimes,
such as~\cite{Ashtekar:2018lag,Ashtekar:2018cay,Ashtekar:2020ckv}, or the unimodular quantization of spherically symmetric spacetime \cite{Gielen:2025ovv}, can in principle provide input for  $\langle \hat \Gamma^\alpha_{\; \mu \nu} \rangle$, to our knowledge they have't done so yet. Thus, meanwhile we will stick to~\cite{Koch:2025yuz}.
The line element
for spherically symmetric spacetime discussed in~\cite{Koch:2025yuz} reads
\be\label{eq_LineElement}
ds^2= n^2(r) g(r) d t^2-\frac{1}{g(r)} dr^2-r^2d\theta^2-r^2\sin^2(\theta)d\phi^2,
\ee
where we opt to use the convention that $c=1$.
This line element  has two unknown functions $n(r)$ and $g(r)$.
In general relativity the solutions for the two functions are
\bea\label{eq_ncl}
n(r)&=&1,\\
\label{eq_gcl}
g(r)&=& 1-  \frac{2 G M}{r}-\frac{r^2 \Lambda }{3}.
\eea
Upon quantization, the metric functions are promoted to field operators.  
For $g(r)$ we get
\bea
g &\rightarrow &\hat g
\eea
with the canonically conjugated momentum $\hat p$. Both operators $\hat g$ and $\hat p$ obey the canonical commutation relations.
The other metric function $g_{00}= n^2 g$, becomes determined through a constraint as
\bea
n&\rightarrow &\hat n
=\frac{\hat p}{r\Gamma}\label{eq_Constraintn},
\eea
where $\Gamma= i c (G/(c \hbar))^{1/4} \gamma /(G M_{\langle\hat g  \rangle})$ is a global integration factor with units of energy which we parameterized in terms of the gravitational couplings $G$, the central gravitational mass $M_{\langle\hat g  \rangle}$ and a dimensionless parameter $\gamma$.
The Hamiltonian of this quantum system in the Einstein-Hilbert truncation is given by
\be\label{eq_H}
\hat H=\frac{1-r^2\Lambda }{r}\hat p-\frac{(\hat g \hat p)}{r},
\ee
where we remember that observable operator products such as the Hamiltonian are Weyl ordered.
For the purpose of q--desics, we are interested in expectation values of metric operators $\hat A$. These can be obtained from the Hamiltonian (\ref{eq_H}) by virtue of the Heisenberg relation 
\be\label{eq_Ehrenfest}
i \hbar\frac{d \langle \hat A\rangle }{dr}= +
\langle [\hat A,\hat H] \rangle +i \hbar \partial_r \langle \hat A\rangle.
\ee
Solving this first order differential equation for different integer valued positive powers $a$ and $b$ of the canonical variables $\hat g$ and $\hat p$ yields the
expectation values
\be\label{eq_VevGeneral}
\langle (\hat g^a \hat p^b)\rangle=
\sum_{i,j,k\ge 0}^{i+j+k=a} r^{b-j+2k}
C_{j,b}
\left(\begin{array}{ccc}
    & a&  \\
     i&j&k 
\end{array}\right)
\left( \frac{(-\Lambda) }{3}\right)^k.
\ee
Each integration gives one additional integration constant, which we label $C_{j,b}$. These constants  characterize the wavefunction of quantum-spacetime $| \Psi\rangle$ in terms of the eventually measurable different averages (\ref{eq_VevGeneral}). 
The constants are almost completely free (depending on $| \Psi\rangle$), taking into consideration that they have to conspire such that 
they obey quantum uncertainty relations, as discussed in~\cite{Koch:2025yuz}.
Note that we have only proven (\ref{eq_VevGeneral}) up to power $a,b\le 5$. This is sufficient for all following results except of $\langle \hat g'/\hat g\rangle$. Thus, our result for $\langle \hat g'/\hat g\rangle$ is a systematic uncertainty of higher order operators.

\subsection{Selected Expectation Values}\label{sec:VEVS}

As we will see, in static spherically symmetric vacuum regions, the q--desic equation (\ref{eq_Geodesic1}) contains six types operator-expectation values. These are ``measurable observables'' in the sense that by measuring the associated displacements of test particles, we can gain information on these expectation values of hermitian quantum operators. In contrast, there are many other quantities such as for example $\langle \hat g \rangle$ which are entitled to carry the name ``observables'' in the sense that they are expectation values of an hermitian quantum operator
(Note that $\langle \hat g \rangle\neq \langle 1/\hat g \rangle^{-1}$). However,
such quantities are not measurable in the sense, that we can not obtain direct information on them by performing measurements of geodesics, or other sorts of displacements. 

Thus, our first task is to evaluate relevant expectation values that appear in the q--desic equation and in the definition of null curves. For this,  we can use the result given in  (\ref{eq_VevGeneral})
\begin{enumerate}
    \item $\langle \hat 1 \rangle$:\\
For this operator we find
    \bea\label{eq_1vev}
\langle \hat  1 \rangle&=& (1+\epsilon_{0,0}).
\eea
If we could impose the normalization of the spacetime wavefunction as in quantum mechanics of a single particle, we could set $\epsilon_{0,0}$ to zero.
\item $\langle \hat g \rangle$:\\
    The expectation value of the metric function is given by
    \bea\label{eq_gvev}
\langle \hat g \rangle&=&(1+\epsilon_{0,0})+\frac{C_{1,0}}{r}-\frac{r^2 \Lambda(1+\epsilon_{0,0})}{3}.
\eea
Clearly, this expectation value has the functional form of the classical solution (\ref{eq_gcl}).
Thus, we reparametrize the integration constant $C_{1,0}$,
in terms of the classical mass $M_0$,
eventually shifted by the dimensionless parameter $\epsilon_{1,0}$
\be\label{eq_C10}
C_{1,0}=-2G M_0 (1+\epsilon_{1,0}).
\ee
In the following, we will use the definition (\ref{eq_C10}), but keep in mind that $\epsilon_{1,0}$  just represents one of the many integration constants of the spherically symmetric quantum spacetime. To make this clear,  this constant was equipped with a subindex to remind us of the origin of this constant. 
This contribution is absent from the classical approach. The quantum modification resulting from the integration constant can be interpreted as a variation in the theory's parameters.
%
    \item $\langle \hat g \hat n^2 \hat g' \rangle$:\\
By using the Heisenberg relation and the Hamiltonian (\ref{eq_H}) we can write the derivative of the metric operator as a simple function of the metric operator
\be\label{eq_gdot}
\hat g'=\frac{1}{i \hbar}[\hat g, \hat H]=\frac{1-\Lambda r^2- \hat g}{r}.
\ee
We further can use the constraint (\ref{eq_Constraintn}) to express the operator in terms of the canonical operator $\hat p$. Thus, we find that
\be
\langle \hat g \hat n^2 \hat g' \rangle=
\frac{\langle \hat g \hat p^2 (1-\Lambda r^2- \hat g) \rangle}{r^3 \Gamma^2}
\ee
Now, that the expectation value is written as sum of  of Weyl ordered products of  canonical operators we can evaluate it by using the relation (\ref{eq_VevGeneral})
\be\label{eq_gn2gp}
\langle \hat g \hat n^2 \hat g' \rangle=
\frac{2 r^4 \Lambda(r^2 \Lambda -3)C_{0,2}-3r (3 + r^2 \Lambda)C_{1,2}-9 C_{2,2}}{9r^3 \Gamma^2}.
\ee
    \item 
    $\langle \hat 1/\hat g \rangle$:\\
    The VEV of this inverse operator was found in~\cite{Koch:2025yuz}. We define 
    \be
    \label{eq:inverse_lng}
    \ln(\hat g)= - \sum_{n=1}^\infty\frac{\hat U^n}{n}
    \ee
    with $\hat U =(1-\hat g)$. With these it follows
    \bea
    \left \langle\frac{1}{\hat g} \right \rangle = \left \langle\frac{\tilde \delta \ln (\hat g)}{\tilde \delta \hat g} \right \rangle  &=&
    \sum_{n=1}^\infty
     \left \langle
     \hat U^{n-1}
     \right \rangle\\ &=& 3-3\langle \hat g\rangle+\langle \hat g^2\rangle+ {\mathcal{O}}\left(\left \langle
     \hat U^3
     \right \rangle\right),\label{eq_ginv1PN}
    \eea   
    where we expanded to next to leading order of the small operator $\hat U$.
    \item 
    $\langle \hat g'/\hat g \rangle$:\\
The expectation value of this ratio of operators was derived in~\cite{Koch:2025yuz}. Acting with radial derivative on eq.~(\ref{eq:inverse_lng}) we obtain

    \bea
    \label{eq_gprime/g_vev}
    \langle \hat g'/\hat g \rangle =
    \sum_{n=1}^\infty 
    \langle \hat U^{n-1} \hat g' \rangle
    = \sum_{n=1}^\infty 
    \langle(1 - \hat g)^{n-1} \frac{(1- r^2 \Lambda - \hat g )}{r} \rangle =  \langle \frac{1- r^2 \Lambda - \hat g }{r} \rangle + \mathcal{O}(\langle \hat U^1 \rangle) \approx \frac{1- r^2 \Lambda }{r} - \frac{\langle \hat g \rangle}{r}. 
    \eea
    
	Inserting the expression for $\langle \hat g \rangle$ given in eq.~\ref{eq_gvev}, we obtain,
    \bea
   \langle \hat g'/\hat g \rangle = \frac{1- r^2 \Lambda }{r} - \frac{1}{r} \left( (1+\epsilon_{0,0})+\frac{C_{1,0}}{r}-\frac{r^2 \Lambda(1+\epsilon_{0,0})}{3} \right)  + \mathcal{O}(\langle \hat U^1 \rangle)
    \eea
    where we expanded to leading power of the small operator $\hat U$.
    \item $\langle \hat n'/\hat n \rangle$ and $\langle \hat g^2 \hat n \hat n' \rangle$:\\
    First, we need to write the operator $\hat n'$ in terms of the canonical operators $\hat g$ and $\hat p$. Using the chain rule on the constraint~(\ref{eq_Constraintn}) we obtain
    \be
    \hat n'=\frac{\hat p'}{r \Gamma}-\frac{\hat p}{r^2 \Gamma}.
    \ee
    By applying the Heisenberg equation to the momentum operator ($i\hbar \hat p'=[\hat p, \hat H]=i \hbar \hat p /r$) we find that
\be\label{eq_ndot00}
\hat n '=0.
\ee
    This implies that 
    all
    expectation values which are proportional to $\hat n'$ vanish
    \be\label{eq_ndot0}
\langle \hat n'/\hat n \rangle = \langle \hat g^2 \hat n \hat n' \rangle=0.
    \ee
\item $\langle \hat n^2 \hat g\rangle $ and $\langle \hat n^2 \hat g^2\rangle$:\\
Although these expectation values do not enter the q--desic equation~(\ref{eq_GeodSym}) directly, we will nevertheless need them in our discussion of null curves.
 It is straight forward to calculate that
\be\label{eq_fvev}
\langle \hat n^2 \hat g\rangle=\frac{C_{0,2}}{\Gamma^2}+\frac{C_{1,2}}{r \Gamma^2}-\frac{r^2 \Lambda C_{0,2}}{3 \Gamma^2}.
\ee
This agrees with the classic metric expectation value $g_{00}$, if we define
\bea\label{eq_C02}
C_{0,2}&=&\Gamma^2(1+\epsilon_{0,2}),\\ \label{eq_C12}
C_{1,2}&=&-2 G M_0 \Gamma^2 (1+\epsilon_{1,2}),
\eea
and subsequently send $\epsilon_{0,2}$ and $\epsilon_{1,2}$ to zero.
Note that the relations (\ref{eq_C02}) and (\ref{eq_C12}) are just  redefinitions of integration constants called $\epsilon_{i,j}$ instead of $C_{i,j}$. 
Another useful identity is given by
\be\label{eq_gp}
\hat g'=\frac{1-\Lambda r^2- \hat g }{r}.
\ee
The final expectation value we will need is 
\be\label{eq_n2g2vev}
\langle \hat n^2 \hat g^2\rangle  r^2 \Gamma^2= r^2 C_{0,2}-\frac{2}{3}r^4 \Lambda C_{0,2}+ \frac{1}{9} r^6 \Lambda^2 C_{0,2}+2 r C_{1,2} -\frac{2}{3} r^3 \Lambda C_{1,2} + C_{2,2}.
\ee
\end{enumerate}
The VEVs presented in this subsection are parametrized by the integration constants $C_{i,j}$ which encode both the classical aspects and the quantum nature of the gravitational wave function $|\Psi\rangle$.
For some of these constants, we were able to rewrite the constants, such that it is split into a classical part and a quantum part~(\ref{eq_C10}, \ref{eq_C02}, \ref{eq_C12}). Even though, the symbol $\epsilon_{i,j}$ suggests that the quantum part is a small correction with respect to the classical part, this does not have to be the case and for certain spacetimes the quantum part may even dominate over the classical part $\epsilon_{i,j}\ge 1$.

\subsection{Mass-coherent States}

In the previous subsection we redefined integration constants (\ref{eq_C10}), (\ref{eq_C02}), and (\ref{eq_C12}) that arose from different expectation values. These constants allow to draw complementary conclusions on the mass $M_0$ at $r=0$.
Depending on the spacetime state $|\Psi\rangle$, the numerical values of these constants may or may not be equal. 
We will call a state  ``mass-coherent''  $|\Psi_M\rangle$ when all possible definitions of mass that arise from different observables, agree upon each other.
For example, we can consider  
$g_{00}$ (\ref{eq_C12}) and the
inverse of $g_{11}$ (\ref{eq_C10}). if
\be\label{eq_MassCoherent}
M_0(1+\epsilon_{1,0})
=M_0(1+\epsilon_{1,2})= M_0, \quad \Leftrightarrow \quad\epsilon_{1,0}=\epsilon_{1,2}=0
\ee
then $|\Psi\rangle=|\Psi_M\rangle$ is a mass-coherent state of the operators $\hat n^2 \hat g$ and $\hat g$.
In this context we want to note that it may well be that it is not possible to achieve complete mass-coherence for all observables mentioned in the previous subsection. The reason for this is that most of the corresponding operators do not commute, e.g.
\be\label{eq_nonCom}
\left [(\hat n^2 \hat g), \hat g \right]\neq 0.
\ee
This implies that the corresponding integration constants (including $\epsilon_{1,0}$ and $ \epsilon_{1,2}$) have to obey inequalities arising from the uncertainty relations~\cite{Koch:2025yuz}.
We leave the study of a potential interplay between coherent states and uncertainty relations to a future project and return to the topic of this paper, the motion of test particles.

\section{q--desic Motion in SER}

Now, that we have obtained the most relevant expectation values of metric operators in quantum gravity using the SER approach, we can proceed to use them in~(\ref{eq_Geodesic1}).
%
\subsection{q--desic Equation in Spherical Background}

For the quantum spacetime described above,
the q--desic equation (\ref{eq_Geodesic1}) reads
\be\label{eq_GeodSym}
\left(
\begin{array}{c}
   \dot  u^t  \\
   \dot  u^r\\
   \dot  u^\theta\\
   \dot  u^\phi
\end{array}
\right)
=
\left(
\begin{array}{c}
-\langle\hat n^2\hat g'/(\hat n^2\hat g)\rangle u^r u^t-2 \langle\hat n'/\hat n\rangle u^r u^t\\
(u^r)^2 \langle\hat g'/(2 \hat g)\rangle+\left\langle\hat g(r (u^\theta)^2+r \sin^2\theta (u^\phi)^2-  \hat n^2 \hat g' (u^t)^2/2 )\right\rangle
-\langle \hat g^2 \hat n \hat n'  \rangle (u^t)^2\\
 2 u^r u^\theta /r - \cos \theta \sin \theta (u^\phi)^2\\
- 2(u^r u^\phi+ r \cot \theta u^\theta u^\phi)/r
\end{array}
\right).
\ee
Without loss of generality, we define $\theta=\pi/2$ and study motion in the equatorial plane. Thus, (\ref{eq_GeodSym}) simplifies to 
\be\label{eq_GeodSym2}
\left(
\begin{array}{c}
   \dot  u^t  \\
   \dot  u^r\\
   \dot  u^\theta\\
   \dot  u^\phi
\end{array}
\right)
=
\left(
\begin{array}{c}
-\left\langle
\frac{\hat n^2\hat g'}{\hat n^2\hat g}\right\rangle u^r u^t\\
\left\langle
\frac{\hat g'}{2 \hat g}
\right\rangle (u^r)^2 +r \left\langle\hat g\right\rangle (u^\phi)^2- 
\left\langle \hat n^2 \hat g\hat g'\right\rangle
 \frac{(u^t)^2}{2} \\
0\\
 -\frac{2}{r} u^r u^\phi
\end{array}
\right),
\ee
where we used the identity (\ref{eq_ndot0}).
As mentioned before, for all operator products, it is implicit that we are using Weyl ordering of the operators as they appear on the right hand side.
Note that in a Weyl ordered product including
operators such as $\hat n$ and the corresponding
inverse operators  $\hat n^{-1}$, it is not
guaranteed that operators will always directly meet with their inverse. Therefore, we can not simply cancel the $\hat n^2$ in the numerator and in the denominator of the first line of (\ref{eq_GeodSym2}).
To simplify this problem, it is sometimes convenient to work with conserved quantities.

\subsection{Energy and Angular Momentum}

For some cases, the geodesic equation (\ref{eq_GeodSym2}) can be further simplified by the use of conserved quantities, which correspond to integration constants of symmetries of the problem. 
The operator valued Lagrangian (\ref{eq_SRPP1b}) for the problem in the equatorial plane reads
\be\label{eq_hatLSER}
\hat L = \hat n^2 \hat g \; (u^t)^2- \frac{1}{\hat g}(u^r)^2 -r^2 (u^\phi)^2.
\ee
This Lagrangian provides two conserved momenta,  which in asymptotically flat spacetime are called energy $\hat E$ 
and angular momentum operator $\hat l$
\bea
\label{eq_EnOp}
\hat E= \hat n^2 \hat g \; u^t & {\text{with}} &\quad\frac{d}{dt} \hat E=0, \quad \Rightarrow \quad E=\langle  \hat n^2 \hat g\rangle \; u^t  \\
\hat l = r^2 u^\phi
 & {\text{with}}&\quad \frac{d}{dt} \hat l=0 \quad \Rightarrow \quad l=  r^2 u^\phi.
 \label{eq_AngMomOp}
\eea
As consistency check, let's recover the first and the fourth component of (\ref{eq_GeodSym}) from the conserved quantities (\ref{eq_EnOp} and \ref{eq_AngMomOp}).
Taking the derivative of (\ref{eq_AngMomOp}) with respect to the proper time and solving for $\dot u^\phi$, we get
\be
\dot u^\phi= -\frac{2}{r} u^r  u^\phi,
\ee
which is identical to the fourth component of the geodesic equation.
Analogously, when we take the derivative of the energy relation (\ref{eq_EnOp}) and solve for $\dot u^t$.  After using (\ref{eq_ndot0}) and taking the VEV we get
\be
\dot u^t=  -\left\langle\frac{\hat n^2\hat g'}{\hat n^2\hat g} \right \rangle u^r u^t,
\ee
which coincides with the first component of (\ref{eq_GeodSym2}).
The q--desic equation (\ref{eq_GeodSym2}) has many potential applications.
We will analyze 
complementary examples consisting of horizons in radial motion and circular trajectories of massive particles.

\subsection{Horizons for Radial Motion}
\label{subsec_Horizon}

We start by remembering the definition of an event horizon in the context of classical geodesics, before revisiting this concept in the light of the q--desics framework.
A classical null curve is characterized
by the condition $ds^2=0$. 
For radial motion and the line element (\ref{eq_LineElement}), this means that
\be\label{eq_ds0}
ds^2=0 \quad \Rightarrow \quad dr^2\frac{1}{g(r)}=n^2 g(r)dt^2.
\ee
We can solve this relation for the radial velocity
\be\label{eq_vr}
(v^r)^2\equiv \left(\frac{d r}{dt}\right)^2= n^2 g^2(r).
\ee
The horizon radius $r_H$ can be defined as the radius of vanishing radial velocity $v^r=0$. The  solutions of this classical condition are the zeros of the function $g(r)$, since classically $n$ becomes a non-vanishing constant. The zeros of $g$ are found at two 
radii, one known as the cosmological horizon radius (for $\{r,\, 1/\sqrt{\Lambda}\}\gg GM_0$)
\be\label{eq_rC}
r_C\approx \sqrt{\frac{3}{\Lambda}}
\ee
and one as the Schwarzschild radius for $r\ll r_C$
\be\label{eq_rS}
r_S\approx2 G M.
\ee

Now, we whish to implement the condition for null curves (\ref{eq_ds0}) in the context of the quantum operators $\hat g$ and $\hat n$.
When doing this, we 
recur to the corresponding relation in terms of expctation values
(\ref{eq_properTimeu2}).
After choosing the option of a null-curve and assuming radial motion we get
\be\label{eq_vr2}
(u^r)^2\equiv
\left(\frac{dr}{dt}\right)^2=\frac{\langle \hat n^2 \hat g \rangle}{\langle \hat g^{-1} \rangle}.
\ee
To finds the horizon(s), we need to know the zeros of this expression.
As long as the denominator can be expanded in small powers of $\hat U$, 
and $\langle \hat g^{-1} \rangle$ does not diverge, it can not contribute to the
zeros of (\ref{eq_vr2}). Thus,
the zeros of (\ref{eq_vr2}) are given by
the zeros of the 
the numerator $\langle \hat n^2 \hat g \rangle$. Assuming again $\{r,1/\sqrt{\Lambda}\}\gg GM_0$, we find that the cosmological horizon remains unchanged, while for $r\ll 1/\sqrt{\Lambda}$ the Schwarzschild horizon may experience 
modifications
\bea
r_{QC}&=& \sqrt{\frac{3}{\Lambda}},\\
r_{QS}&=& 2 G M \frac{1+\epsilon_{1,2}}{1+\epsilon_{0,2}}.
\eea
As can be seen, when $\epsilon_{1,2} > \epsilon_{0,2}$ the quantum Schwarzschild horizon increases relative to the classical value, whereas for $\epsilon_{1,2} < \epsilon_{0,2}$ the horizon decreases.

%
The classical horizon (\ref{eq_rC}) is recovered for 
\be\label{eq_CorHor}
\epsilon_{0,2}=\epsilon_{1,2}.
\ee
Note that the horizon condition (\ref{eq_CorHor}) is complementary but not identical to the condition we found in the context of mass-coherence (\ref{eq_MassCoherent}).

\subsection{Circular Motion of Massive Particles}
\label{subsec_circular}

Classical circular motion of massive particles provides a unique orbital velocity for a given radial distance $r$.
To derive it, we consider the $r$ component of (\ref{eq_GeodSym2}), replacing every operator by its expectation value and setting $u^r=0=\dot u^r$
\be\label{eq_GodClcirc}
\left(-\frac{G M}{r^{2}} + \frac{r \Lambda  }{3}\right) u^{t}(\tau)^{2}
\;+\; r \, u^{\phi}(\tau)^{2}=0.
\ee
Now, we can express $u^t$ in terms of the classical conserved energy (\ref{eq_EnOp}) and complement it with 
the classical version of (\ref{eq_properTimeu2}),  written in terms of the energy
\be\label{eq_E2cl}
E^2=
-\frac{\bigl(6\,G\,M - 3\,r + r^{3}\,\Lambda  \bigr)
       \!\left(1 + r^{2}\,u^{\phi}(\tau)^{2}\right)}
      {3\,r}.
\ee
Inserting (\ref{eq_E2cl}) into (\ref{eq_GodClcirc}) we obtain a relation for the proper angular velocity
\be\label{eq_uphicl}
(u^\phi)^2=\frac{-3\,G\,M \;+\; r^{3}\,\Lambda  }
     {3\,(3\,G\,M - r)\,r^{2}}.
\ee
Now we define the orbital velocity $v$ with respect to the coordinate time of a static observer at radius $r$ as
\be\label{eq_v2def}
v^2=r^2 (u^\phi)^2\frac{(n^2 g)}{E^2}
\ee
and insert (\ref{eq_uphicl}) and (\ref{eq_E2cl}) into (\ref{eq_v2def}).
The classical orbital velocity then reads
\be\label{eq_v2cl}
v^2=\frac{G\,M}{r}\;-\;\frac{r^{2}\,\Lambda  }{3}.
\ee
The relations (\ref{eq_uphicl}, \ref{eq_v2cl}) are important for observational purposes, since they
can be translated directly to Kepler's Third Law.

Let's now explore the corresponding circular motion for q--desics by setting $\dot u^r=0=u^r$ in the second component of 
(\ref{eq_GeodSym}). We get
\be\label{eq_RadMot1}
\frac{(u^t)^2}{2} \langle \hat g \hat n^2 \hat g'\rangle=
r \langle\hat g \rangle (u^\phi)^2.
\ee
Next, we replace the velocities $u^t$ and  $u^\phi$ by the use of the conserved quantities   (\ref{eq_EnOp}) and (\ref{eq_AngMomOp})
and we invoke the expectation values listed in subsection (\ref{sec:VEVS}). This leads to
\be\label{eq_GeodSym22}
\frac{l^2}{3 r^3}(3r-6 G M (1+\epsilon_{1,0})+ r^3 \Lambda)
=E^2\frac{\left(
-18 G^2 M^2 (1+\epsilon_{2,2})+r^4 (1+\epsilon_{0,2})\Lambda (-3+ r^2 \Lambda)+ 3 G M r (1+\epsilon_{1,2})(3+r^2 \Lambda)
\right)
}{ 6 G M (1+\epsilon_{1,2})-r(1+\epsilon_{0,2})(3-r^2 \Lambda)}.
\ee
In this relation the particle's angular momentum $l$ and it's energy $E$ still appear as independent parameters.
These parameters can be related further by solving (\ref{eq_properTimeu2}) for $E$ and inserting the result into (\ref{eq_GeodSym2}).
After straight forward manipulations 
we find the proper angular velocity
%
\be\label{eq_uphi2circ}
\small
(u^\phi)^2=
\frac{
      \frac{1}{3r^{2}}\left(-18\,GM\bigl(1+\epsilon_{2,2}\bigr)
      +\frac{r^{4}}{GM}\bigl(1+\epsilon_{0,2}\bigr)\,\Lambda\bigl(-3+r^{2}\Lambda\bigr)
      +3\,r\bigl(1+\epsilon_{1,2}\bigr)\bigl(3+r^{2}\Lambda\bigr)\right)}
      {%
          6\,GM\!\left(
                3+2\epsilon_{1,2}
                 +2\epsilon_{1,0}\bigl(1+\epsilon_{1,2}\bigr)
                 +\epsilon_{2,2}\right)
          +\frac{r^2}{GM}\bigl(1+\epsilon_{0,2}\bigr)\bigl(3-r^{2}\Lambda\bigr)
          -r\!\left(
                15+6\epsilon_{1,0}+9\epsilon_{1,2}
               -3r^{2}\Lambda
               -2r^{2}\epsilon_{1,0}\Lambda
               -r^{2}\epsilon_{1,2}\Lambda
               +2\epsilon_{0,2}\bigl(1+\epsilon_{1,0}\bigr)\bigl(3-r^{2}\Lambda\bigr)
             \right)
      }
\ee
\normalsize
We generalize the definition (\ref{eq_v2def}) to the q--desic case
\be\label{eq_v2defQ}
v^2=r^2 (u^\phi)^2\frac{\langle\hat n^2 \hat g\rangle}{E^2}=\frac{r^2 (u^\phi)^2}{1+r^2 (u^\phi)^2}
\ee
and use (\ref{eq_uphi2circ}) 
together with the energy. This allows to write (\ref{eq_v2defQ}) as
\be\label{eq_v2circ}
v^2=\frac{  -18\,G^{2}M^{2}\bigl(1+\epsilon_{2,2}\bigr)
      - r^{4}\bigl(1+\epsilon_{0,2}\bigr)\,\Lambda\!\left(3 - r^{2}\Lambda\right)
      + 3\,GMr\bigl(1+\epsilon_{1,2}\bigr)\!\left(3 + r^{2}\Lambda\right)
     }
     {\left[ 
     \,6\,G M\,\bigl(1+\epsilon_{1,0}\bigr) - r(1+\epsilon_{0,0})\left(3 - r^{2}\Lambda\right)\right]
     \left[
     \,6\,G M\,\bigl(1+\epsilon_{1,2}\bigr) - r(1+\epsilon_{0,2})\left(3 - r^{2}\Lambda\right)\right]
     }.
\ee
The orbital velocity (\ref{eq_v2circ}) is the q--desic generalization of the classical relation (\ref{eq_v2cl}). 
It contains,
as we will see in the following section, potential dynamical aspects which can not be captured in terms of the usual geodesics in classical spherically symmetric spacetime.

\section{Discussion}

We now turn to additional aspects of q--desics and the preceding results.

\subsection{Averages and Operator Equations}

A recurring theme in quantum gravity and quantum field theory is whether one should first take the expectation value of an operator and then insert it into some dynamical equation, or rather first use the full operator equation of motion and only afterward take its expectation value. 
The two procedures need not yield identical results. 
In particular, simply taking a classical or bare quantum equation of motion for a c-number variable and then replacing that variable by its expectation value often fails to capture genuine quantum back-reaction or fluctuation effects. 
By contrast, starting from the operator-level equation of motion and then taking the expectation value incorporates commutators and uncertainty relations in a nontrivial way, hence retaining richer quantum information.

A particularly illuminating example is the well-known Ehrenfest theorem in standard (non-relativistic) quantum mechanics, which describes how the expectation values of position and momentum evolve in time. 
Consider a single freely moving particle of mass $m$ in a one-dimensional potential $V(x)$. 
Its Heisenberg operator equations of motion yield, upon taking expectation values, the Ehrenfest relations:
\begin{equation}\label{eq:ehrenfest1}
m\,\frac{d}{dt}\langle x\rangle \;=\;\langle p\rangle,
\quad\quad
\frac{d}{dt}\langle p\rangle \;=\;-\bigl\langle V'(x)\bigr\rangle.
\end{equation}
The first equation is reminiscent of the classical relation $p = m\,\dot{x}$. 
However, the second equation will only reduce to Newton's second law,
\[
\frac{d}{dt}\langle p\rangle \;=\; -\,V'(\langle x\rangle),
\]
if $\langle V'(x)\rangle$ happens to be equal to $V'(\langle x\rangle)$. 
In general, these two expressions differ because
\[
\langle V'(x)\rangle \;=\; V'(\langle x\rangle) 
\quad\text{would require}\quad 
V'\bigl(x\bigr) \;\text{to be linear in}\; x,
\]
or at least to be a function for which the expectation value and the function evaluation commute. 
For a generic potential, such as a cubic $V(x)$ whose derivative $V'(x)$ is quadratic, one encounters the familiar discrepancy between $\langle x^2\rangle$ and $\langle x\rangle^2$. 
Hence, identifying $\langle x\rangle,\langle p\rangle$ with classical position and momentum and then applying the classical force law is just an approximation.

\noindent
The key takeaway is that, in order to derive proper equations of motion for expectation values, one must preserve the full operator structure for as long as possible, only transitioning to expectation values at the very end. This ensures that crucial quantum corrections which are stemming from commutators, fluctuations, and nonlinearities—are retained rather than being prematurely discarded.

\subsection{Comments on the q--desic derivation and interpretation}
\label{subsec_Comments}

A few comments on the derivation and interpretation of the q--desic equation  are in place here. 

\begin{itemize}
    \item Metric operators do not necessarily commute. Thus, there will be ordering ambiguities in the definition of (\ref{eq_GammaDef}). In this work we addressed this issue by imposing Weyl ordering, whenever products of the metric operators appear in observables.
\item 
Another crucial ingredient in the derivations of the q--desic equation is the assumption of a well-defined inverse metric operator (\ref{eq_ginv}). This operator is used when deriving the q--desic equation. Whether the inverse metric operator can be consistently formulated, or whether this is at least challenging, depends on the approach. In some approaches, such as perturbative QG~\cite{Donoghue:1994dn}, string theory~\cite{Aharony:1999ti}, stochastic gravity~\cite{Verdaguer:2006iy}, or static equal radius quantization~\cite{Koch:2025yuz}, this operator is consistently defined. In many other approaches such as causal set theory~\cite{Bombelli:1987aa}, or those based on canonical quantization (for a review see~\cite{Thiemann:2022fwv}), a consistent definition of inverse metric operators is not a fundamental concept or poses a major challenge. Thus, there is currently no consensus on whether this operator can be consistently well-defined in QG or not. Moreover, depending on the QG scenario, the absence of an inverse metric operator can go hand in hand with violations of Lorentz invariance or CPT symmetry~\cite{Kostelecky:2003fs,kostelecky2021searches}. However, since there is some evidence that both Lorentz invariance and CPT symmetry still hold in the Planck regime~\cite{Knorr:2018fdu,Eichhorn:2025ilu}, one might suspect that the same is true regarding the existence of an inverse metric operator, even if the mathematical formulation is challenging.

In any case, the q--desic equation can only be derived and consistently solved in the context of QG approaches that allow for a well-defined inverse metric operator.
\item 
 The q--desic  (\ref{eq_Geodesic1}) is more general than the well-known g--eodesic equation, but it
 reduces to the latter whenever we can approximate averages of products by products of averages. We can quantify this in terms of the covariance 
\bea\label{eq_CovG}
{\mbox{Cov}}^\beta_{\; \mu \nu}(\Gamma)&=&
\left \langle
\hat \Gamma^\beta_{\; \mu \nu}
\right\rangle- \Gamma^{\beta}_{\; \mu \nu}(\langle g \rangle)
\\ \nonumber
&=&
\left \langle \frac{1}{2}
\left(
\hat g_{\mu \alpha,\nu}+\hat g_{\nu \alpha,\mu} -\hat g_{\mu \nu,\alpha} \right)
\hat g^{\alpha \beta}\right\rangle
- \frac{1}{2}
\left(\left \langle\hat g_{\nu \alpha,\mu} \right\rangle + \left \langle\hat g_{\mu \alpha,\nu} \right\rangle - \left\langle\hat g_{\mu \nu,\alpha}\right\rangle
\right)\left \langle
\hat g^{\alpha \beta}\right\rangle
\\ \nonumber
&= &
\left \langle
\left \{\hat X_{\mu \nu \alpha}, \hat g^{\alpha \beta} \right \}
\right\rangle +\left \langle
\left [\hat X_{\mu \nu \alpha}, \hat g^{\alpha \beta}\right ]
\right\rangle ,
\eea
where we defined
\be
\hat X_{\mu \nu \alpha}\equiv
\frac{1}{4}
\left(
\hat g_{\mu \alpha,\nu}+\hat g_{\nu \alpha,\mu} -\hat g_{\mu \nu,\alpha} \right).
\ee
For metric states with ${\mbox{Cov}}^\beta_{\; \mu \nu}(\Gamma)=0$, geodesics and q--desics are identical. For non-vanishing covariance, 
both equations (and their solutions) are different. 
In our implementation, the commutator part of (\ref{eq_CovG}) vanishes exactly,
due to the afore mentioned Weyl ordering prescription. The remaining anticommuting part
\be
{\mbox{Cov}}^\beta_{\; \mu \nu}(\Gamma)=\left \langle
\left \{\hat X_{\mu \nu \alpha}, \hat g^{\alpha \beta} \right \}
\right\rangle
\ee
is only sensitive to quantum fluctuations of the $\hat X_{\mu \nu \alpha}$ and $\hat g^{\alpha \beta}$ around their mean values.
If these fluctuations are statistically uncorrelated, the covariance vanishes and
q--desic motion and geodesic motion are indistinguishable. This can happen for example if $|\Psi\rangle$ is an eigenstate of either $\hat X_{\mu \nu \alpha}$ or $\hat g^{\alpha \beta}$,
a fact we have already noted when we invoked (\ref{eq_ginv}) to derive the q--desic equation.
\item 
We can attempt to examine the covariance in relation  (\ref{eq_CovG})
for the concrete examples.
Most straightforwardly, we can use the  non-relativistic circular motion (\ref{eq_RadMot1}):
For this case, it
reads 
\be
  {\mbox{Cov}}^1_{\; 00}=\frac{1}{4}\left \langle 
\{ (\hat n^2 \hat g')_W, g \}
\right \rangle-\frac{1}{4}\langle n^2 g \rangle' \langle g \rangle.
\ee
By using the known expectation values we can calculate this expression directly. 
From the result we can read-off explicitly what a vanishing covariance means in terms of the integration constants $\epsilon_{i,j}$, as it will be shown
in subsection \ref{subsec_Obs}.\\
Another example of finding the covariance can be seen in bottom-up approaches such as stochastic gravity~\cite{Verdaguer:2006iy,Hu:2008rga} where the Einstein equations are supplemented by a stochastic source accounting for the fluctuations of quantum fields. In this framework, the metric is decomposed into a background metric and a perturbation metric. The covariance of the metric perturbations is obtained from the difference between the two-point function and the product of expectation values and is separates into intrinsic fluctuations and a induced fluctuations. The intrinsic part is associated with fluctuations of the initial state of the metric, while the induced part is proportional to the noise kernel and thus is connected to fluctuations of quantum fields. In the absence of matter, the stochastic source and hence the induced part vanish, leaving us only with the intrinsic fluctuations. In our work, the metric is promoted to an operator, and we have not yet included matter contributions, which we plan to study the inclusion of matter in future works. It will be particularly interesting to analyze the two-point functions of the metric in the presence of matter fields and compare them with those computed in stochastic gravity, where both intrinsic and induced contributions are present. We will explore this in more detail in future work.
\end{itemize}

\subsection{Relation to the Effective Average Action Approach}

The q--desic framework, as introduced in this work, offers a novel and extended perspective on the apparent structure of spacetime in quantum gravity. It is instructive to compare it with the effective average action (EAA) formalism, which plays a central role in the asymptotic safety program~\cite{Weinberg:1976xy,Wetterich:1992yh,Morris:1993qb,Bonanno:2000ep,Reuter:2001ag,Litim:2002xm,Reuter:2004nv,Bonanno:2006eu,Niedermaier:2006ns,Percacci:2007sz,Dupuis:2020fhh}.

In the EAA approach, the central object is the scale-dependent effective action $\Gamma_k[g_{\mu\nu}]$, which governs the dynamics of the effective (i.e., expectation value) metric field. The EAA satisfies a functional renormalization group (FRG) equation that systematically accounts for quantum fluctuations above a momentum scale $k$. By varying $\Gamma_k$ with respect to the metric, one obtains scale-dependent field equations,
\begin{equation}
\frac{\delta \Gamma_k}{\delta g_{\mu\nu}(x)} = 0,
\end{equation}
whose solutions can be interpreted as providing $\langle \hat{g}_{\mu\nu}(x) \rangle$, the expectation value of the quantum metric operator at scale $k$. In this sense, the EAA formalism provides a dynamical framework for obtaining effective geometries from quantum gravity, including the effects of higher-derivative curvature operators such as $R^2$, $R_{\mu\nu}R^{\mu\nu}$, and others. Therefore, the EAAs used by effective quantum-gravitational theories, such as the asymptotic safety program may give complementary and additional information with respect to the results presented in section \ref{sec_SER}. This can be expected, since the results presented in section \ref{sec_SER} are obtained from canonically quantizing the Einstein-Hilbert action, while neglecting higher-derivative operators such as $R^2$, $R_{\mu\nu}R^{\mu\nu}$, which are explicitly allowed in the EAA approach.

By contrast, the q--desic program is agnostic regarding the origin of the expectation values of geometric quantities; it assumes no specific derivation method—canonical, covariant, or effective-action-based.
As an example for one of these methods, we use the geometric expectation values employed in the q--desic construction by performing a canonical quantization of the Einstein--Hilbert action~\cite{Koch:2025yuz}.
However, it would certainly be interesting to 
investigate whether alternative approaches to QG, such as the ones using the EAA can be used to provide  the explicit formulas for the expectation values $\langle \hat{\Gamma}^\rho_{\mu\nu} \rangle$, which serve as input for the q--desic equations.

\subsection{Relation to other Approaches with Non-Geodesic Motion}

There are well-known theories, which have some common aspects with the q--desics discussed above:

\begin{itemize}
    \item Spinning top:\\
The well-established theory of spinning particle motion in curved spacetime, was developed by Mathisson, Papapetrou, and later refined by Dixon and Hojman~\cite{Mathisson:1937zz,Papapetrou:1951pa,Corinaldesi:1951pb,Hojman:1978yz,Hojman:1978wk,Hojman:2012me,hojman:1975}. In that context, extended test bodies with intrinsic spin follow non-geodesic trajectories due to spin-curvature coupling. The resulting Mathisson-Papapetrou-Dixon (MPD) equations describe how the spin tensor interacts with the Riemann curvature, leading to deviations from geodesic motion that are fully classical in origin. A key feature of the MPD framework is that, in addition to the metric, the particle couples to the spacetime connection via its spin degrees of freedom. In this framework, modifications from the classical geodesics are suppressed by inverse powers of the particle mass, which is the reason why massless spinning tops were only introduced much later in an extended formalism~\cite{Armaza:2016vfz}.

The q--desic approach shares a conceptual similarity in that it also predicts deviations from classical geodesics, but for fundamentally different reasons. While the theories of spinning tops retain a quantum property of the test particle (spin) when formulating equations of motion, the deviations in the q--desic approach arise from the quantum nature of the gravitational background itself. As a result, q--desic trajectories reflect quantum corrections to the affine structure, even for spinless test particles.

Thus, while both frameworks involve non-geodesic motion driven by connection-level information, the source of the deviation differs: classical spin-curvature interaction in the MPD case versus quantum-induced modifications of the connection in the q--desic case. This highlights the role of the affine connection as a carrier of geometric information that can influence particle motion beyond the metric alone.
\item 
Modified Newtonian Dynamics (MOND)\\
Another context in which non-geodesic motion plays a central role is Modified Newtonian Dynamics (MOND), originally proposed by Milgrom to explain galactic rotation curves without invoking dark matter. In MOND and its relativistic extensions (such as TeVeS and MONDian scalar-tensor theories), test particles deviate from geodesics of the background metric due to modifications in the gravitational force law, particularly at accelerations below a critical scale \( a_0 \). These deviations are typically introduced phenomenologically or via additional fields that mediate modified gravity effects~\cite{Milgrom:1983pn,Milgrom:1983zz,Bekenstein:2004ne}.

The q--desic framework also leads to non-geodesic motion, but for conceptually distinct reasons. Rather than modifying gravity at large distances or low accelerations, q--desics arise from the quantum structure of spacetime itself. This allows for deviations from classical geodesics even in regimes where general relativity would traditionally be expected to hold, without invoking new dynamical fields or modifying the underlying force law.

While both approaches predict departures from geodesic motion, MOND does so to account for large-scale astrophysical phenomena, whereas q--desics aim to capture quantum corrections to spacetime geometry in semiclassical gravity. Their similarity, which arises despite these fundamental differences, lies in the central role of the affine connection 
and in the fact that they can induce large-scale deviations from classical geodesics.
\item 
Fifth force theories \\
Scalar–tensor theories replace the fixed Newton constant by a dynamical scalar field that couples non-minimally to the metric, so test bodies feel an additional Yukawa-like “fifth force’’ typically mediated by the scalar~\cite{Damour:1994zq,Copeland:2006wr,Clifton:2011jh,Fujii:2003pa,Gasperini:2001pc,Damour:2002nv,Damour:2002mi,Brax:2010gi,Brax:2011ja,Hinterbichler:2010es,Khoury:2003aq}, or a renormalization scale~\cite{Koch:2010nn,Koch:2014cqa,Koch:2016uso}. Laboratory  setups, planetary ephemerides and other precision probes have been studied to detec, or tightly bound, the strength and range of such forces \cite{Adelberger:2003zx,Lemmel:2015kwa,Koch:2022cta}. 

Within these theories, the motion of test particles in curved spacetime may look distinct from classical GR. However, when this motion is calculated, it is assumed that a test particle follows a geodesic for a given spacetime metric and the difference only arises because the metric itself differs from the one predicted by GR. This is the main difference between these theories and the q--desic concept, where the motion itself is explicitly and not just seemingly non-geodesic.
\end{itemize}


\subsection{Observable Deviations at Large and Small Scales}
\label{subsec_Obs}
\begin{figure}[hbt]
\centering
\includegraphics[width=0.56\linewidth]{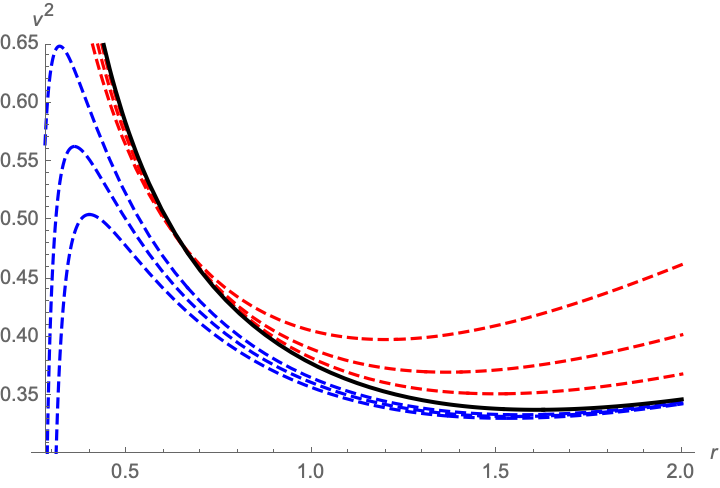}
    \caption{Orbital velocity as a function of the radius. For all curves we used in natural untits $G=1,\; M=0.1,$ and $\Lambda=-0.05 $. The black curve corresponds to the classical relation (\ref{eq_v2cl}). 
    In terms of (\ref{eq_v2circ}) this classical velocity-distance relation corresponds to $\epsilon_{0,0}=\epsilon_{1,0}=\epsilon_{0,1}=\epsilon_{1,2}=\epsilon_{2,2}=0$.
    The colored curves are also based on (\ref{eq_v2circ}), but allowing one single parameter to be different from zero. For the blue curves we chose $\epsilon_{2,2}=\{0.3,\; 0.4,\; 0.5\}$, while for the red curves we we used $\epsilon_{0,0}=\{-0.2,\; -0.4,\; -0.6\}$.}
    \label{fig:vr}
\end{figure}

In subsection~\ref{subsec_Horizon} we showed that the q--desic framework can shift the classical event–horizon radius.  
Turning to circular orbits in subsection~\ref{subsec_circular}, we identified a clear departure from the geodesic approximation: the orbital speed derived from the q--desic equation~(\ref{eq_v2circ}) no longer satisfies the classical relation~(\ref{eq_v2cl}).  
This discrepancy prompts several questions:
\begin{itemize}
    \item ``{\it Can the classical relation for orbital velocities (\ref{eq_v2cl}) be recovered from the q--desic's result
(\ref{eq_v2circ}) e.g. by the choice of 
a special spacetime wave function $|\Psi\rangle=|\Psi_0\rangle$?}''\\
The special spacetime state $|\Psi_0\rangle$ that allows to reduce 
the former to the latter
is characterized by the choice of the integration constants $\epsilon_{i,j}$.
We find that for
\be\label{eq_epsionCirc0}
(\epsilon_{1,0},\, \epsilon_{0,2},\, \epsilon_{1,2}, \,\epsilon_{2,2})=  (0, \, 0, \,0 ,\, 0), 
\ee
the q--desic circular orbits (\ref{eq_v2circ}) become indistinguishable from geodesal orbits (\ref{eq_v2cl}).
Thus, the answer to the first question is: ``{\it Yes, unless some up to now unexplored uncertainty relation forbids to take the limit~(\ref{eq_epsionCirc0})}''.
\item ``{\it What is the leading observational short range difference between (\ref{eq_v2cl}) and (\ref{eq_v2circ}) given that $|\Psi\rangle\neq |\Psi_0\rangle$}?''\\
To answer this question, let's consider orbits that are both far outside from the Schwarzschild horizon (\ref{eq_rS}) and still safely inside from the cosmological horizon (\ref{eq_rC}), such that their radius obeys $2 G M\ll r_o< \sqrt{3/\Lambda}$.
For these orbits we can use the doubly small expansion parameter
\be\label{eq_delta}
\frac{2 G M}{r} \approx  r_o^2 \frac{\Lambda}{3}\equiv \delta\ll 1.
\ee
Using this expansion, we find that for $\epsilon_{1,2}=\epsilon_{0,0}+\epsilon_{0,2}+\epsilon_{0,0}\epsilon_{0,2}$ the Newtonian limit is recovered and that (\ref{eq_v2circ}) can be written as
\bea\label{eq_v2approx}
v^2&=&\frac{G M}{r}
\;-\;
\frac{1}{3(1+\epsilon_{0,0})}\,r^{2}\,\Lambda\\ \nonumber
&&+\frac{2 G^{2} M^{2}}{r^{2}}\frac{1+ 2 \epsilon_{0,0}(1+\epsilon_{0,2})+\epsilon_{0,0}^2(1+\epsilon_{0,2})+\epsilon_{1,0}+\epsilon_{0,2}(2+\epsilon_{1,0})-\epsilon_{2,2}}{(1+\epsilon_{0,0})(1+\epsilon_{0,2})}
-G M r\frac{\Lambda}{3} \frac{1-2\epsilon_{0,0}-\epsilon_{0,0}^2+2\epsilon_{1,0}}{(1+\epsilon_{0,0})^2}
+{\mathcal{O}}\left( \delta^3,\; r^3 \right).
\eea
From this relation we see that the first two terms which are of order $\delta^1$,
 correspond to the classical result (\ref{eq_v2cl}) if we further set $\epsilon_{0,0}\rightarrow 0$. The leading  corrections to this are given by the third and the fourth term and are both of the order $\delta^2$.
The third term corresponds to a short-term correction. Short-range corrections were to be expected. For example, such QG corrections to the Newton potential and other typical quantities were calculated in ~\cite{Ward:2002zw,Faller:2007sy,Donoghue:2012zc,Donoghue:1994dn,Donoghue:1993eb,BjerrumBohr:2002ks}. However, while typical short-range QG corrections are believed to be of order $\sim 1/r^3$, the third term is of different nature, since it already contributes in at order $\sim 1/r^2$. This term is potentially much less suppressed than perturbative estimates made us believe.
Thus, the answer to the second question is: ``{\it The leading short range correction is given by the $\epsilon_{i,j}\neq 0$ in the third term''}
Exploring this difference brings a novel perspective to solar-system tests of GR~\cite{Reynaud:2008yd}, which certainly can 
gain information on the wave function of this system by further constraining the parameters.
\item ``{\it Are there relevant  long range observational differences between (\ref{eq_v2cl}) and (\ref{eq_v2circ}) given that $|\Psi\rangle\neq |\Psi_0\rangle$}?''\\
The leading long-range correction to the velocities of circular orbits is given by the fourth term of (\ref{eq_v2approx}). Depending on the sign and magnitude of the spacetime parameter $\epsilon_{1,0}$, this term can either act attractive or repulsive. We can not know how likely or unlikely this is to happen, but it is our task to 
either constrain or confirm this possibility for different gravitational systems with approximate spherical symmetry.
This type of scenario can be observationally highly relevant:
It is known that constant and linear corrections in $r$ to the squared orbital velocities
can play an important role in our understanding of galaxy rotation curves and the ever-present yet elusive dark matter~\cite{Zasov:2017gem,Carlson:1996js,Rodrigues:2022oyd,Marra:2020sts,Rodrigues:2018duc}  (maybe even variable values of $\Lambda$ may be considered here~\cite{Yin:2024hba,Sola:2017znb,Alvarez:2020xmk,DiValentino:2021izs,Verde:2019ivm}).
One may argue that universal corrections to the $v^2$ formula 
have limited use because they
lack the flexibility of addressing individual features of different galactical systems~\cite{Rodrigues:2020ybx,Rodrigues:2018duc}. However, such arguments do not apply in the context of q--desics,
since each gravitational system and each galaxy has it's own local wave function $| \Psi\rangle$, and thus it's own values for $\epsilon_{i,j}$. So we can conclude that: {\it Yes, for certain spacetime wavefunctions $|\Psi\rangle\neq |\Psi_0\rangle$, there are relevant large scale contributions.}
\item ``{\it Is there also a constant correction to $v^2$?''}\\
Formula (\ref{eq_v2approx}) shows no constant correction term, so the answer to this question seems to be ``{\it No}''.  However, this conclusion is premature. 
If we expand to higher orders in $\delta$ (or relax the normalization condition (\ref{eq_1vev})), we also find a constant term contributing to 
(\ref{eq_v2approx}).
\end{itemize}
To exemplify the tendency of the short- and long-distance corrections,
let's plot the orbital velocity (\ref{eq_v2circ}) as a function of the radial distance from the center, as shown in figure~\ref{fig:vr}.
This figure illustrates that QG gravity contributions can be phenomenologically relevant at short, intermediate, and large distance scales, depending on the structure of the spacetime wave-function $|\Psi\rangle$, which is encoded in the values of the parameters $\epsilon_{i,j}$. The long-term goal is to use model predictions, such as the rotation curves in Figure~\ref{fig:vr}, to constrain or confirm the q--desic parameter space $\epsilon_{i,j}$. This program would be in the spirit of~\cite{Kaloper:2006pj}. However, realistic orbits are neither exactly circular nor are realistic spacetimes exactly spherically symmetric. Thus, for such a phenomenological program, we still need to calculate orbits that go beyond circular motion and eventually relax the assumption of spherical symmetry.

\section{Summary and conclusions}

In this work, we have introduced and developed the concept of q--desics, generalizations of classical geodesics, arising in the context of quantum gravity. These trajectories are defined by replacing the classical Christoffel symbols with the expectation value of the quantum connection operator, $\langle \hat{\Gamma}^\rho_{\mu\nu} \rangle$. This approach extends the traditional reliance on the expectation value of the metric, $\langle \hat{g}_{\mu\nu} \rangle$, and allows for a richer characterization of quantum geometrical effects.

We derived the q--desic equation from both Lagrangian and Hamiltonian perspectives. 
We then used previous results on spherically symmetric static backgrounds within canonical quantum gravity and applied them in the context of the q--desic equation.
We analyzed the resulting equations of motion in complementary regimes, consisting of radial null-curves and  circular orbits, revealing quantum-induced modifications to classical behavior. This case study revealed that, depending on the quantum state of spacetime $|\Psi\rangle$, we can expect to measure, or constrain both short-  and long-range modifications from classical velocity curves~(\ref{eq_v2approx}).
We have also discussed conceptual differences and parallels between q--desics and other instances of non-geodesic motion, such as those appearing in the motion of spinning particles (Mathisson-Papapetrou-Dixon theory) in MOND-like gravitational theories, or fifth force theories. 

Looking ahead, q--desics  open several avenues for further exploration. They provide a framework for probing quantum corrections to spacetime beyond the effective metric approximation, and may be extended to incorporate more general quantum states, and alternative quantum gravity models. Most importantly, q--desics give a completely new tool for observational searches for QG effects at large distances.\\

\textbf{Note added:}
After the article's first appearance on the arXiv, we became aware of three related approaches:
\begin{enumerate}
    \item The impact of virtual gravitons and backreaction was studied in \cite{Dalvit:1997yc,Dalvit:1999wd}. The calculations use perturbation theory, expanding the metric around a semiclassical background $\tilde g_{\mu \nu}=g_{\mu \nu}+\kappa h_{\mu \nu}$, 
    followed by path--integral methods to derive an effective action for both the background and a test particle. Although the techniques and language differ from our approach, it is natural to assume that this effective test--particle action corresponds to the VEV of our operator--valued action (\ref{eq_SH}), i.e. $\langle \hat S_H \rangle$. Extremizing this expectation value of the average action yields the VEV of equation (\ref{eq_deltax2}), leads to
    \[
    \frac{d^2 x^\mu}{d\lambda^2}
    \;+\;\Gamma^\mu_{\alpha\beta}\bigl(\langle  \tilde g_{\gamma \delta}\rangle\bigr)
    \,\frac{dx^\alpha}{d\lambda}\,\frac{dx^\beta}{d\lambda}
    \;+\frac{1}{m} \langle  \tilde g^{\mu \nu}\rangle \frac{\delta \Delta S_m}{\delta x^\nu}=\;0\,,
    \]
    where $\Delta S_m$ contains 
    the $\sim h_{\mu \nu}^2$ contributions to the point particle action. Even if $\Delta S_m=0$, this is not the classical geodesic equation, since the effective metric $\tilde  g_{\mu \nu}$ contains contributions from a nonclassical spacetime state. Nevertheless, 
    it contains less information than the q--desic equation (\ref{eq_Geodesic1}), 
    due to the subtlety explained between equations (\ref{eq_geodOp}) and (\ref{eq_Geodesic1}), 
    where the VEV is taken before acting with the inverse metric operator.
    The difference with respect to the q--desic is exactly the information about the spacetime state, which can not be obtained in terms of universal loop calculations and can not be expressed in terms of the universal constants ($G,\, \hbar, \, c, \, M$).
        \item Perturbative quantum fluctuations of the metric field have also been considered in the definition of geodesic motion with respect to the orthonormal frame of an observers worldline~\cite{Parikh:2020kfh,Chawla:2021lop,Bak:2022oyn,Bak:2023wwo,Cho:2021gvg,Hsiang:2024qou}. It was shown that a path integral over such fluctuations produces a characteristic but $\Psi$-state dependent stochastic force $N_\Psi$, modifying the geodesic equations. For example, even in the Newtonian limit 
    \be
     \ddot{ \vec x}= - \vec \nabla \phi + N_\Psi \vec x + \dots\quad .
    \ee
    Such an effect could in principle be identified by observing the correlations of numerous different geodesics. In contrast, the q--desic  (\ref{eq_Geodesic1}), can already be manifest in a single trajectory.
    \item The definition of a spatial distance operator in the context of QG provides the starting point for another series of papers~\cite{Piazza:2025uxm,Nitti:2024iyj,Piazza:2022amf,Piazza:2021ojr}. The authors do not attempt to compute a semiclassical metric; instead they focus on the spacetime state $\Psi$. They show that, under certain assumptions, the state is not peaked around the classical value and distances become nonadditive. In our approach, this nonadditivity arises as a special case of the covariance in equation (\ref{eq_CovG}). In this sense, those works anticipated several of our findings based on this non-trivial covariance.
\end{enumerate}
Interestingly, in approaches~1 and 2 the deviations from classical geodesics are expected only at extremely short distances, whereas in approach~3 the dominant deviations occur at very large distances. The q--desics derived here allow for both short-- and long--range modifications; see Fig.~\ref{fig:vr}.

\section*{Acknowledgements}
The authors acknowledge TU Wien Bibliothek for financial support through its Open Access Funding Programme.
\'Angel Rinc\'on thanks Silesian University in Opava for all their support during the development of this research work.

\newpage
%

\bibliography{ref.bib}

\end{document}